\documentclass{aa}
\usepackage{graphics}
\usepackage{psfig}

\newcommand{\hi}{H\,{\sc i}}
\newcommand{\hii}{H\,{\sc ii}}
\newcommand{\halpha}{H\,$\alpha$}

\newcommand{\etal}{et al.}

\newcommand{\Ros}{{\em ROSAT}}

\newcommand{\eg}{e.g. }
\newcommand{\ie}{i.e. }
\newcommand{\etc}{etc.}
\newcommand{\arcs}{\hbox{$^\prime$}\hskip -0.1em\hbox{$^\prime$}}
\def\arcm{\hbox{$^\prime$}}

\begin{document}

   \title{X-ray emission from the Sculptor galaxy NGC~300}
   \author{A.M.~Read \and W.~Pietsch}
   \offprints{A.M.~Read, e-mail address aread@mpe.mpg.de}
   \institute{Max-Planck-Institut f\"ur extraterrestrische Physik,
              Gie\ss enbachstra\ss e, D--85748 Garching, Germany}
   \date{Received date; accepted date}
   \markboth{A.M.~Read \& W.~Pietsch: The X-ray source population within NGC~300}{}

\abstract{
We report here the results of a full analysis of all the \Ros\ PSPC spectral imaging
observations and all the \Ros\ HRI high resolution imaging observations 
of the very nearby ($D=2.1$\,Mpc) Sculptor galaxy, NGC~300. Many point sources
are detected within the field, several of them showing evidence for variability, and we
present full source lists detailing their X-ray properties, and attempt to classify them
on the basis of their temporal, spectral and multi-wavelength characteristics. A black
hole X-ray binary candidate, a supersoft source and several supernova remnants and \hii\
regions are detected in X-rays, as is unresolved, possibly diffuse emission, accounting
for perhaps $\sim$20\% of the total NGC~300 X-ray (0.1$-$2.4\,keV) luminosity
($5.8\times10^{38}$\,erg s$^{-1}$). We compare the X-ray source luminosity distribution
of NGC~300 with that of other nearby galaxies, and we also compare NGC~300 with its
Sculptor neighbours, concluding that it is a quite an unremarkable system, showing no
unusual X-ray (or other multi-wavelength) properties. It may be one of the best examples 
of a completely typical normal quiescent late-type spiral galaxy.
\keywords{Galaxies: clusters: Sculptor -- Galaxies: individual: NGC~300 -- 
Galaxies: ISM -- Galaxies: spiral -- X-rays: galaxies}
}

   \maketitle

\section{Introduction}
\label{sec_intr}

NGC~300, a near face-on SA(s)d galaxy, is one of the five well-known galaxies
making up the nearby Sculptor galaxy group, and appears in fact to be possibly
involved in weak gravitational interaction with one of the other Sculptor
galaxies, namely NGC~55. Though estimates of its distance vary, we here adopt the
value of 2.1\,Mpc used by several authors (e.g. Freedman \etal\ 1992, Blair \&
Long 1997 [hereafter BL97], C\^{o}t\'{e} \etal\ 1997). At this distance, 1\arcm\ corresponds to
0.61\,kpc, and the 
scale of the galaxy, as indicated by its major (D$_{25}$) and minor
axis (de Vaucouleurs \etal\ 1991), is approximately 13.3$\times$9.4\,kpc 
(hence the system is physically quite small). Many
giant \hii\ regions are visible within the galaxy, and there is evidence for many
episodes of star formation having taken place. Interestingly though, it appears
that star-forming activity in the centre of NGC~300 has been suppressed for the
past 10$^{9}$\,yr with respect to the disk, the central arcminute containing only
a modest population of stars with ages less than 1\,Gyr (Davidge 1998). Though the
spiral arms are not so clearly defined, the spiral pattern is very evident (see
e.g. Sandage \& Bedke 1988). NGC~300's close proximity, face-on nature and 
the fact the galactic hydrogen column density in the direction of NGC~300 is 
relatively low (3.6$\times10^{20}$\,atoms cm$^{-2}$; Dickey \& Lockman 1990), thus 
minimizing any absorption of soft X-rays, make it an ideal target for the study 
of X-ray source populations in nearby spiral galaxies. Furthermore, its 
similarity to other small spiral galaxies, notably M33 (see BL97; Haberl \& 
Pietsch 1999) and 
its Sculptor neighbours, is useful in terms of comparisons. 

NGC~300 was first observed in X-rays by the Position Sensitive Proportional Counter
(PSPC) (Pfeffermann \etal\ 1986) on board \Ros\ (Tr\"{u}mper 1983). The \Ros\
X-ray telescope (XRT), with the PSPC at its focal plane, offers three very
important improvements over previous X-ray imaging instruments. Firstly, the
spatial resolution is very much improved, the 90\% enclosed energy radius at
1\,keV being $27''$ (Hasinger \etal\ 1992). Secondly, the PSPC's spectral
resolution is very much better ($\Delta E/E \sim 0.4$ FWHM at 1\,keV) than earlier
X-ray imaging instruments, allowing the derivation of characteristic source and
diffuse emission temperatures. Lastly, the PSPC internal background is very low
($\sim3\times10^{-5}$\,ct s$^{-1}$ arcmin$^{-2}$; Snowden \etal\ 1994), thus
allowing the mapping of low surface brightness emission.
The \Ros\ High Resolution Imager (HRI) on the other hand, because of its excellent
spatial resolution (more like $5''$) and relative insensitivity to diffuse
emission, is an ideal instrument for further investigation into the point source populations. 

A presentation of some of the early NGC~300 PSPC observations has already
been given by Read \etal\ (1997), hereafter RPS97, as part of a homogeneous
analysis of archival \Ros\ data from several nearby spiral galaxies. Several
point sources are seen together with a small amount of unresolved, possibly
diffuse emission. Of the point sources seen, several appear to lie close to
the positions of clusters of stars, and a few appear to coincide with \hii\
regions and Wolf-Rayet stars. The same PSPC observations were also very briefly
discussed by Zang \etal\ (1997). 

In terms of other significant multi-wavelength studies of NGC~300, two which
are important are the SNR identification work of Blair \& Long (1997; BL97)
and Pannuti \etal\ (2000), hereafter P00. P00 made use of both the NGC~300
ROSAT PSPC X-ray results from RPS97 (though not correcting for the
different assumed NGC~300 distances in RPS97 and P00), and the optical work
of BL97, such that several SNR candidates are now thought to exist within
NGC~300. These are discussed throughout the present paper.

Here we report the results of the full \Ros\ observations (46\,ks of PSPC,
40\,ks of HRI) of the field surrounding NGC~300, concentrating primarily on
the point source population within and around the galaxy. The plan of the
paper is as follows. Section\,\ref{sec_obse} describes the observations and the
preliminary data reduction methods used and Sect.\,\ref{sec_resu} discusses
the results as regards the point sources. Section\,\ref{sec_disc} discusses the
X-ray properties of NGC~300, with regard both to its membership of the
Sculptor group and to how it compares to spiral galaxies in general. Finally a
summary is presented in Sect.\,\ref{sec_summ}.

\section{ROSAT observations and preliminary analysis}
\label{sec_obse}

\subsection{PSPC}

\begin{figure*}[thpb]
  \resizebox{14cm}{!}{
    \psfig{figure=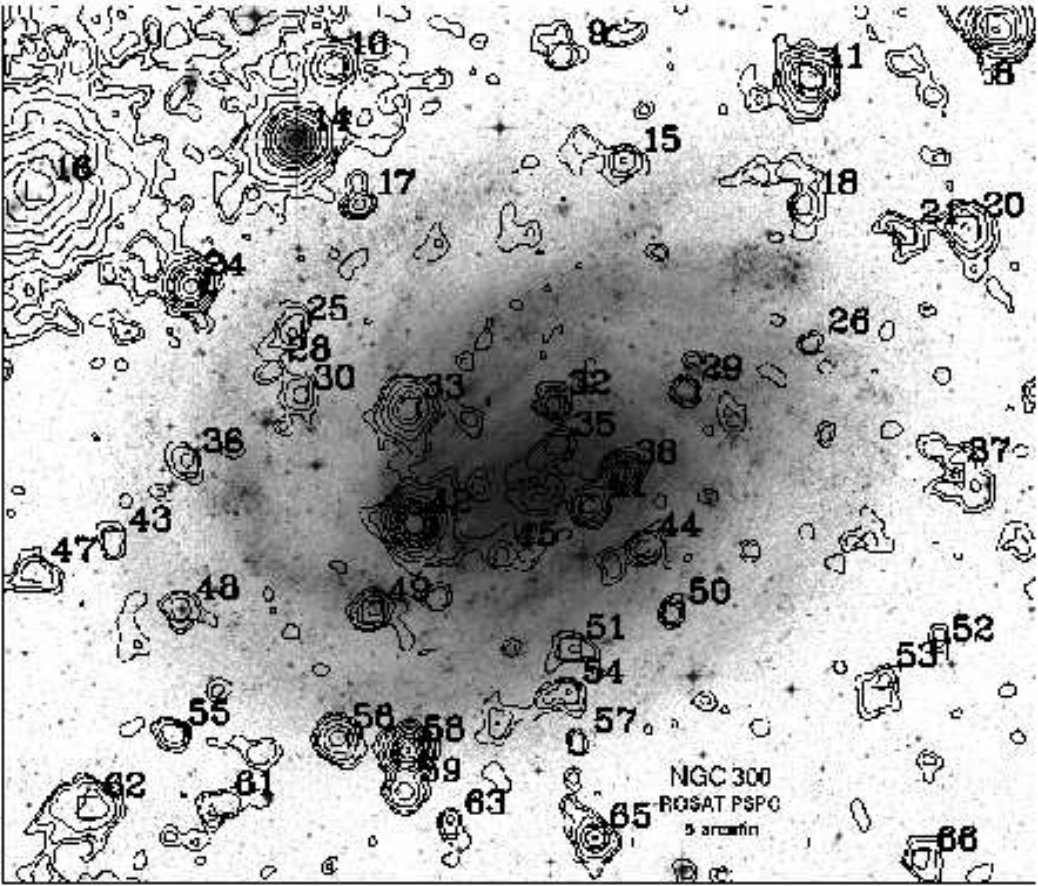,width=14cm,clip=}}
\hfill \parbox[b]{3.5cm}
{\caption{
\Ros\ PSPC contour map of the NGC~300 field in the broad (channels
11$-$235,corresponding approximately to 0.11$-$2.35\,keV) energy band, overlayed on a
Digital Sky Survey (DSS2) red image. The contour levels are at 2, 3, 5, 9, 15, 31, 63,
127, 255, 511 and 1023 $\sigma$ above the background ($\sigma$ being
$2.2\times10^{-4}$\,cts s$^{-1}$ arcmin$^{-2}$, the background level being
$1.2\times10^{-3}$\,cts s$^{-1}$ arcmin$^{-2}$). PSPC source positions, as given in
Table~\ref{table_Psrc}, are marked.
}}
\end{figure*}

A summary log of the entire \Ros\ PSPC and HRI observations of 
the NGC~300 field is given in Table~\ref{table_obse}. 
Because the PSPC is some three times as sensitive as the HRI, we expect,  
from the relevant exposure times of the observations, 
many more counts from the PSPC data. We have analysed all the data extensively, 
and find this to be so. Nevertheless the HRI data are in themselves of 
great interest, and their analysis and subsequent results are described
later. 

NGC~300 was observed with the ROSAT PSPC twice (see Table~\ref{table_obse}). 
Though each PSPC dataset was seen to be very clean, times of 
both very high and very low accepted
event rates and master veto rates were removed. Source detection and
position determination were performed over the full field of view with the EXSAS
local detect, map detect, and maximum likelihood algorithms (Zimmermann et al.
1994) using images of pixel size 15\arcsec. The two eventsets were then shifted
with respect to the prominent X-ray and optical bright star in the field, HD5403,
correcting for the proper motion of the star (Perryman \etal\ 1997) at the epoch
of the ROSAT observations. The two cleaned and position-corrected datasets were
then merged together, and the source detection procedures were re-ran.

\begin{table}
\caption[]{
Summary log of the entire \Ros\ PSPC and HRI observations of
the NGC~300 field. The listed exposure is the nominal 
exposure time. All observations have a nominal pointing 
of 00$^{\rm h}$54$^{\rm
m}$52.0$^{\rm s}$ -37$^{\circ}$41\arcm24.0\arcsec. 
}
\label{table_obse}
\begin{tabular}{lcrcc}
\hline
\noalign{\smallskip}
OBS-ID & Instr. & Exposure & \multicolumn{2}{c}{Observation} \\ 
       &        & Time (s) & Start date & End Date           \\ 
\noalign{\medskip}
(1)    & (2)    & (3)      & (4)        & (5)                \\ 
\noalign{\smallskip}
\hline
\noalign{\smallskip}
600025p-0 & PSPC & 9324 & 28/11/91 & 02/01/92 \\
600025p-1 & PSPC &36693 & 26/05/92 & 29/06/92 \\
600621h   & HRI  &15224 & 08/06/94 & 25/06/94 \\
600621h-1 & HRI  &19138 & 27/05/95 & 28/05/95 \\
600933h   & HRI  & 5619 & 02/06/97 & 03/06/97 \\
\noalign{\smallskip}
\hline
\end{tabular}
\end{table}

Sources accepted as PSPC detections were those with a likelihood L $\geq$10. 
Probabilities P, are
related to maximum likelihood values L, by the relation P$=1-e^{-\mbox{L}}$.
Thus a likelihood L of 10 corresponds to a Gaussian significance of
4.0$\sigma$ (Cruddace \etal\ 1988; Zimmermann \etal\ 1994).

\begin{figure}[thpb] 
\unitlength1.0cm 
\begin{picture}(8.,7.)
\psfig{figure=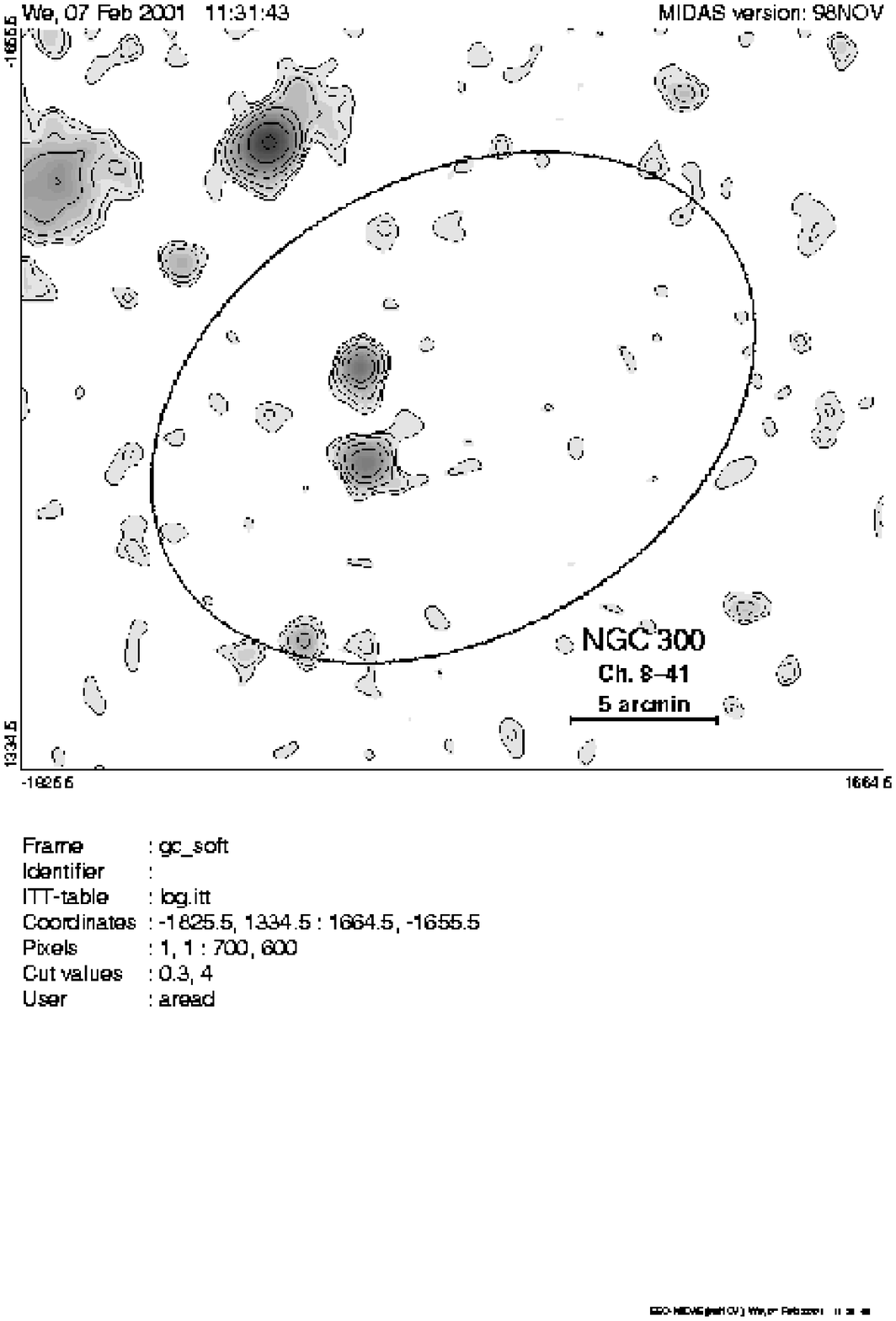,width=8.cm,clip=}
\end{picture} 
\begin{picture}(8.,7.)
\psfig{figure=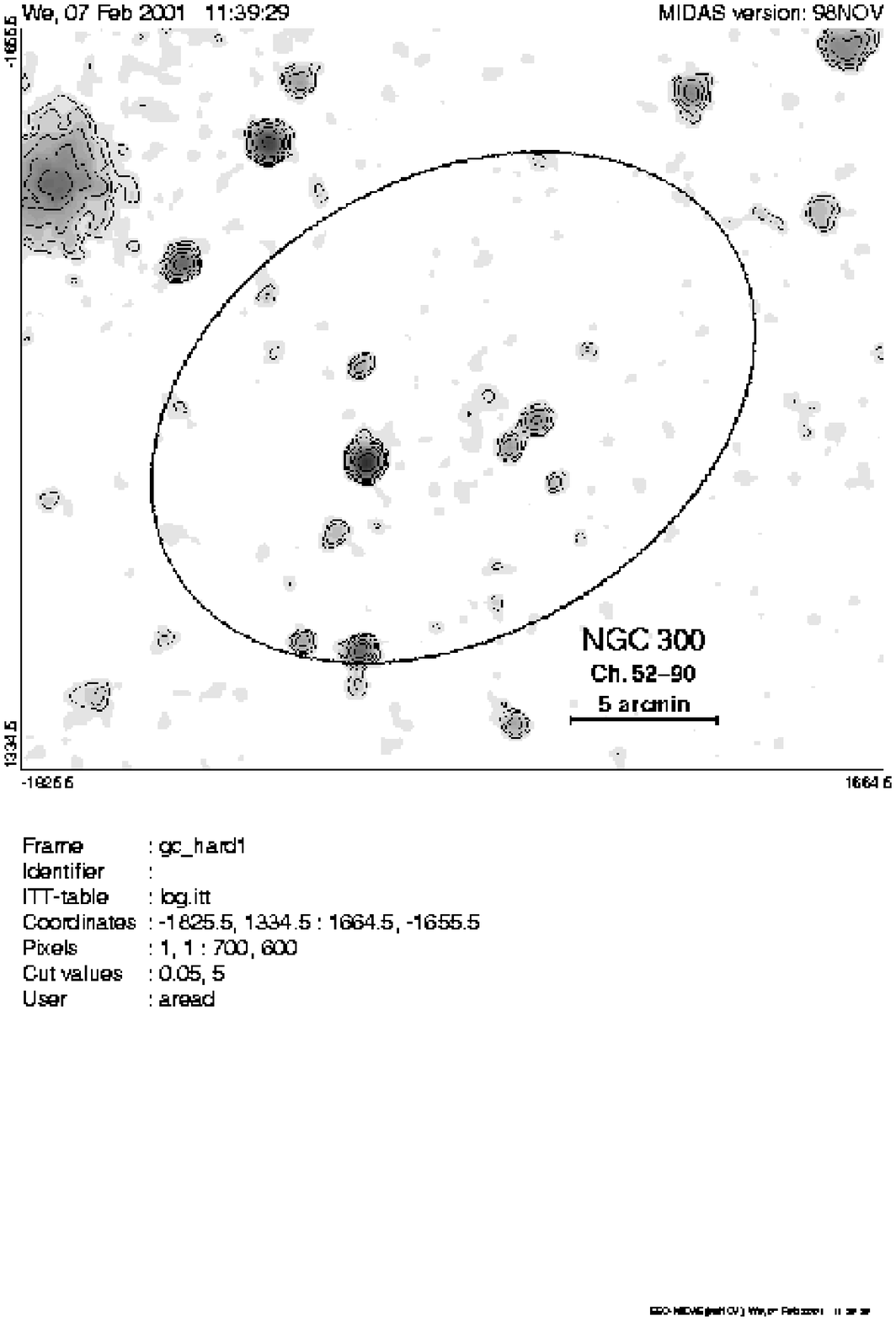,width=8.cm,clip=}
\end{picture} 
\begin{picture}(8.,7.)
\psfig{figure=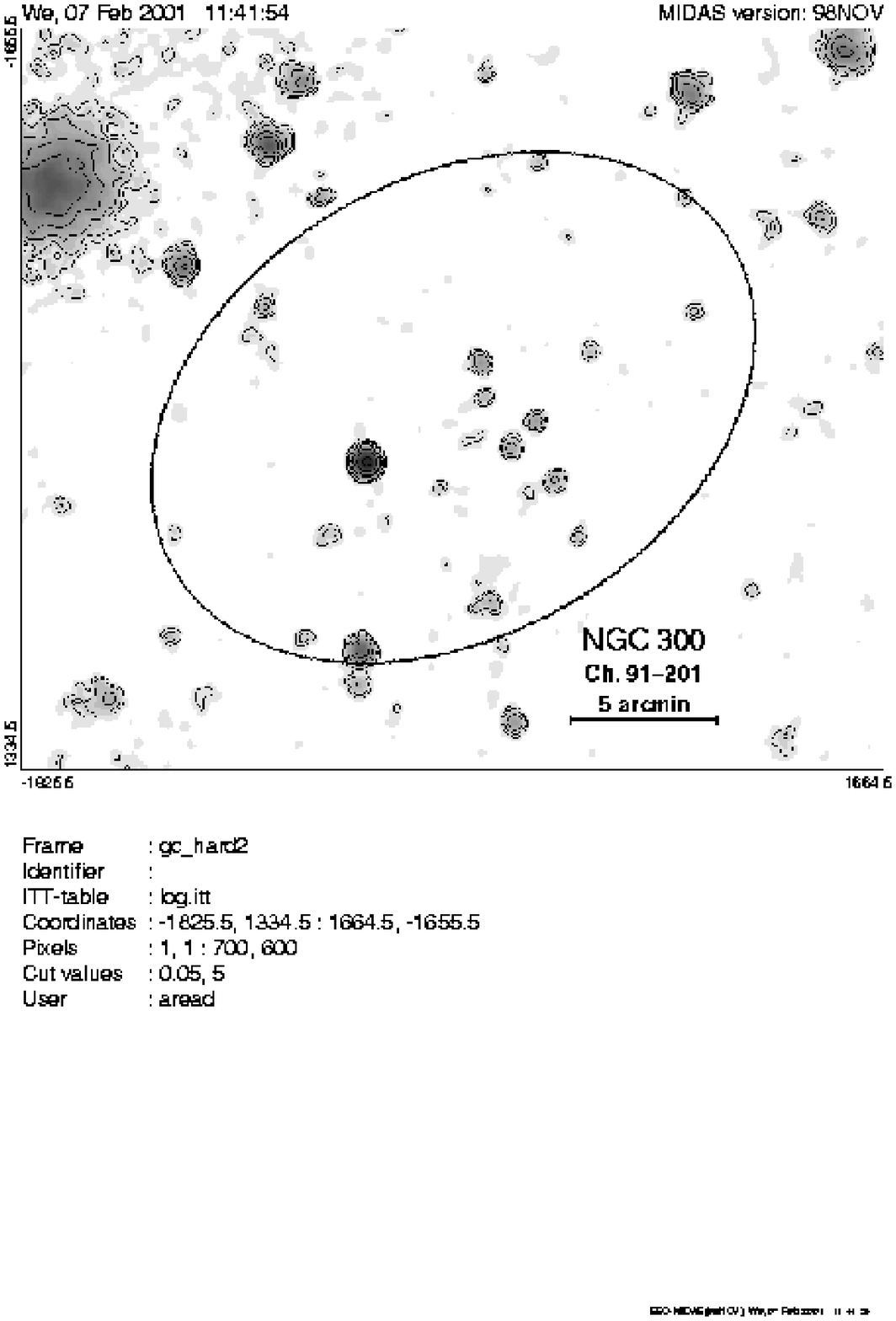,width=8.cm,clip=}
\end{picture} 
\hfill \parbox[b]{8.5cm} 
{\caption{\Ros\ PSPC maps of the NGC~300 field in the (top) soft
(ch.\,11$-$41), (middle) hard~1 (ch.\,52$-$90) and (bottom) hard~2
(ch.\,91$-$201) bands. For the soft band, the contour levels are at 2, 3, 5,
9, 15, 31 and 63 $\sigma$ above the background ($\sigma$ being
$1.4\times10^{-4}$\,cts s$^{-1}$ arcmin$^{-2}$, the background level being
$9.5\times10^{-4}$\,cts s$^{-1}$ arcmin$^{-2}$). For the non-background-limited
hard~1 and hard~2 cases, the contour levels are at 2, 3, 5, 9, 15, 31 and 63
times a value of $3.0\times10^{-4}$\,cts s$^{-1}$ arcmin$^{-2}$. The D25
ellipse of the galaxy is also marked.
}}
\end{figure}

In the following, we concentrate primarily on the sources found within the
optical confines of NGC~300, though we discuss a few other interesting sources
detected close by in Sect.~\ref{sec_res2} and in Appendix.~\ref{sec_appe}.

Figure\,1 shows a broad band (channels 11$-$235, corresponding approximately to
0.11$-$2.35\,keV) contour image of the central $\sim$29\arcm$\times$25\arcm\ region, 
overlayed on a
Digital Sky Survey (DSS2) red image (note the bright optical and X-ray star HD5403, to
the north east, used to register the X-ray coordinates).

Shown in Fig.\,2 are equivalent contour 
images in the soft (ch.\,11$-$41), hard1 (ch.\,52$-$90) and hard2 (ch.\,91$-$201) \Ros\
PSPC bands, with the D25 ellipse of the galaxy marked. 
Within the area covered by Fig.\,1, 47 sources are detected, their source
numbers marked in the figure, 26 of which lie within (or $<1$\arcmin\ from) the optical
disk of NGC~300 (as indicated by the D25 ellipse in Fig.\,2). 
These 26 sources are listed in
Table~\ref{table_Psrc} as follows: source number (Col.\,1), corrected right ascension
and declination (Cols.\,2,\,3), error on the source position (Col.\,4, including a
3\farcs9 systematic attitude solution error), likelihood of existence (Col.\,5), net
broad band counts and error (Col.\,6), and count rates and errors after applying
deadtime and vignetting corrections (Col.\,7). Two hardness ratios are given in Cols.\,8
\& 9, HR1, defined as (hard$-$soft)/(hard$+$soft) (hard and soft being the net counts in
the hard (channels 52$-$201) and soft (channels 11$-$41) bands, respectively), and HR2,
defined as (hard2$-$hard1)/(hard2$+$hard1) (hard1 and hard2 being the net counts in the
hard1 (channels 52$-$90) and hard2 (channels 91$-$201) bands, respectively). The
corresponding errors, as per Ciliegi \etal\ 1997, are also given. While HR1 is most
sensitive to variations in the absorbing column, HR2 traces more the power law index or
temperature. Hardness ratios can be used to give very crude estimates of the spectral
parameters that best describe the source photons (compare \eg the tabulated values with
plots showing the variation of HR1 and HR2 for simple spectral models in \eg Pietsch
\etal\ 1998). The 0.1$-$2.4\,keV flux and X-ray luminosity, assuming a 5\,keV thermal
bremsstrahlung model and a source distance of 2.1\,Mpc (\ie to NGC~300), are given in
Cols.\,10 \& 11. Count rates of the PSPC-detected point sources can be converted into
fluxes, assuming a variety of spectral models. A 1\,keV thermal bremsstrahlung model for
instance, gives rise to fluxes 6\% greater than those given in Table~\ref{table_Psrc}.
Finally note that none of these 26 sources are observed to be significantly
extended. A similar table for the PSPC sources in Fig.\,1 detected {\em outside} of the
D25 ellipse is given in Table~\ref{table_PsrcA} in the appendix (\ref{sec_appe}).

\begin{table*}
\caption[]{
X-ray properties of PSPC-detected point sources within and close to 
the optical confines of NGC~300 (see text).
}
\label{table_Psrc}
\begin{tabular}{lrrrrrrrrrr}
\hline
\noalign{\smallskip}
Src.& \multicolumn{2}{c}{R.A.\,(J2000) Dec} & R$_{err}$ & Lik. & Net counts & Ct.\,rate &
\multicolumn{2}{c}{Hardness ratios} & $F_{\rm x}$ & $L_{\rm x}$ \\
 & ($^{\rm h~~~m~~~s}$) & ($^{\circ}~~~\arcm~~~\arcsec$) & ($\arcsec$) &  & &
(ks$^{-1}$) & (HR1) & (HR2) & ($\frac{10^{-14}{\rm erg}}{{\rm cm}^{2} {\rm s}}$) & 
($\frac{10^{36}{\rm erg}}{{\rm s}}$) \\
\noalign{\medskip}
(1) & (2) & (3) & (4) & (5) & (6) & (7) & (8) & (9) & (10) & (11) \\
\noalign{\smallskip}
\hline
\noalign{\smallskip}
P15 & 00 54 40.62 & -37 32 02.6 &  13.3& 32.3   &53.0(10.5)  & 1.3(3) & 0.36(29) & 0.13(21) &  2.3(0.5) &  11.9 \\
P17 & 00 55 17.90 & -37 33 15.6 &  11.7& 38.4   &46.8(9.9)   & 1.1(2) & 1.00(00) & 0.40(19) &  2.0(0.4) &  10.5 \\
P18 & 00 54 15.60 & -37 33 15.6 &  15.2& 19.0   &40.6(10.3)  & 1.0(3) & 0.36(54) & 0.17(27) &  1.7(0.4) &   9.2 \\
P25 & 00 55 27.34 & -37 36 48.1 &  14.3& 37.3   &55.0(10.7)  & 1.3(3) & 0.68(30) & 0.38(18) &  2.3(0.5) &  12.3 \\
P26 & 00 54 14.01 & -37 37 07.0 &  14.8& 19.8   &19.7(8.2)   & 0.5(2) & 1.00(00) & 0.92(22) &  0.8(0.3) &   4.4 \\
P28 & 00 55 29.96 & -37 37 55.9 &  20.8& 13.9   &20.3(8.7)   & 0.5(2) & 1.00(00) & 0.77(26) &  0.9(0.4) &   4.5 \\
P29 & 00 54 31.90 & -37 38 26.4 &  10.9& 32.4   &39.7(9.4)   & 0.9(2) & 1.00(00) & 0.15(21) &  1.7(0.4) &   8.7 \\
P30 & 00 55 26.41 & -37 38 29.2 &  19.9& 18.7   &37.9(10.6)  & 0.9(3) & 0.84(45) & 0.51(16) &  1.6(0.4) &   8.4 \\
P32 & 00 54 50.56 & -37 38 51.5 &   8.9& 71.8   &57.4(10.2)  & 1.3(2) & 1.00(00) & 0.75(15) &  2.4(0.4) &  12.5 \\
P33 & 00 55 11.01 & -37 38 59.3 &   7.3& 219.3  &237.5(18.3) & 5.6(4) &-0.73(05) &-1.00(00) &  9.9(0.8) &  52.1 \\
P35 & 00 54 49.82 & -37 40 01.0 &  11.8& 27.5   &40.0(9.4)   & 0.9(2) & 1.00(00) &-0.12(21) &  1.7(0.4) &   8.7 \\
P36 & 00 55 42.08 & -37 40 29.5 &  17.2& 20.1   &36.8(10.1)  & 0.9(2) & 0.64(75) &-0.26(23) &  1.6(0.4) &   8.3 \\
P38 & 00 54 41.10 & -37 40 48.9 &   6.7& 158.7  &116.6(13.3) & 2.7(3) & 0.87(14) &-0.33(10) &  4.8(0.5) &  25.4 \\
P41 & 00 54 45.41 & -37 41 43.6 &   7.8& 105.7  &96.4(12.5)  & 2.3(3) & 0.90(19) &-0.12(13) &  4.0(0.5) &  21.0 \\
P42 & 00 55 10.13 & -37 42 13.1 &   4.3& 2458.7 &1015.8(33.5)&24.0(8) & 0.52(03) & 0.06(04) & 42.2(1.4) & 222.5 \\
P44 & 00 54 37.82 & -37 42 51.5 &   9.3& 56.9   &56.9(10.5)  & 1.3(2) & 1.00(00) & 0.24(16) &  2.4(0.4) &  12.4 \\
P45 & 00 54 57.98 & -37 43 06.7 &  15.9& 10.8   &29.4(9.0)   & 0.7(2) & 0.26(57) & 0.46(33) &  1.2(0.4) &   6.4 \\
P48 & 00 55 42.85 & -37 44 36.7 &  16.5& 25.1   &51.6(10.9)  & 1.3(3) & 0.25(27) &-0.03(23) &  2.2(0.5) &  11.7 \\
P49 & 00 55 16.01 & -37 44 39.3 &  12.1& 49.1   &62.9(11.4)  & 1.5(3) & 0.76(23) &-0.16(16) &  2.6(0.5) &  13.9 \\
P50 & 00 54 33.71 & -37 44 44.2 &  12.2& 23.8   &34.8(8.9)   & 0.8(2) & 0.86(55) & 0.32(24) &  1.4(0.4) &   7.6 \\
P51 & 00 54 48.28 & -37 45 43.7 &  21.4& 12.5   &28.2(9.4)   & 0.7(2) & 1.00(00) & 0.48(21) &  1.1(0.4) &   6.2 \\
P54 & 00 54 48.49 & -37 46 58.0 &  11.7& 38.5   &43.9(9.7)   & 1.0(2) & 0.97(41) & 0.64(13) &  1.8(0.4) &   9.7 \\
P56 & 00 55 20.89 & -37 48 14.0 &   8.5& 91.9   &127.1(14.2) & 3.1(3) &-0.18(11) &-0.36(15) &  5.4(0.6) &  28.5 \\
P57 & 00 54 46.90 & -37 48 21.2 &  17.0& 11.1   &20.2(7.8)   & 0.5(2) & 1.00(00) & 0.49(32) &  0.8(0.3) &   4.5 \\
P58 & 00 55 11.08 & -37 48 35.5 &   5.4& 618.1  &309.5(19.6) & 7.5(5) & 0.82(08) & 0.01(07) & 13.1(0.8) &  69.2 \\
P59 & 00 55 11.51 & -37 49 44.2 &  13.1& 37.0   &71.6(11.9)  & 1.7(3) & 0.57(23) & 0.20(17) &  3.0(0.5) &  16.1 \\
\noalign{\smallskip}
\hline
\end{tabular}
\end{table*}

\subsection{HRI}

As discussed earlier, we expected the total PSPC observation to be far more
sensitive than the total HRI observation. Nevertheless we reduced the entire
HRI data collected at three epochs after the PSPC observations (see
Table~\ref{table_obse}) to check for time variability of sources, resolve
confused sources and to obtain improved source positions.

For the analysis we screened the observations for good time intervals longer
than 10\,s; we made no further selections on low background times (which would
have reduced the accepted time by 32\%) as we were mainly interested in
point-like sources and for this purpose, we were still photon limited.
Before merging the datasets for source detection, we checked the attitude
solution of the individual observations using three relatively bright sources
detected in all 3 observations (H3, H13, and H16 of Fig.\,3). The positions of
these sources in the individual pointings agree to better than $\pm 1$\arcsec,
and we therefore did not correct the attitude of the pointings before merging
the data. As with the PSPC, source detection and position determination was
then performed over the full field of view on the merged dataset with the
EXSAS local detect, map detect, and maximum likelihood algorithms (Zimmermann
\etal\ 1994), using images of pixel size 5\arcsec, restricted to HRI raw
channels 1--8 to reduce the detector background. We accepted sources with a
likelihood $\ge 8$ as HRI detections.  The resulting source list was then
shifted with respect to the star HD5403, as done to improve the PSPC field
positioning.

Figure\,3 shows a full, 39.7\,ks HRI contour image (channels 1$-$8) of the
central $\sim$29\arcm$\times$25\arcm\ region (\ie equivalent in area to
Figs.\,1 \& 2). Within the area covered by Fig.\,3, 18 HRI sources are
detected (with a likelihood $L$ $>$8), their source numbers marked in the
figure, 10 of which lie within the optical disk of NGC~300 (as again indicated
by the D25 ellipse). Crosses in Fig.\,3 mark the positions of PSPC-detected 
sources (Fig.\,1). 
In Table~\ref{table_Hsrc}, similarly to
Table~\ref{table_Psrc}, we list these 10 HRI sources as follows: source
number (Col.\,1), corrected right ascension and declination (Cols.\,2,\,3),
error on the source position (Col.\,4, including a 3\farcs9 systematic
attitude solution error), likelihood of existence (Col.\,5), net broad band
counts and error (Col.\,6), and count rates and errors after applying deadtime
and vignetting corrections (Col.\,7). The 0.1$-$2.4\,keV flux assuming again a
5\,keV thermal bremsstrahlung model and Galactic absorption is given
(Col.\,8), as is the X-ray luminosity (Col.\,8), assuming an NGC~300 source
distance of 2.1\,Mpc. Finally, an identification column is given (Col.\,10), listing
which PSPC source
(if any) is a likely counterpart. 
Finally note that only one of the 18 sources is observed to be 
significantly extended (H4 at a likelihood of 82.9; see Sect.~\ref{sec_res2}). 
A similar table for the HRI sources in Fig.\,3 detected {\em outside} of the
D25 ellipse is given in Table~\ref{table_HsrcA} in Appendix~\ref{sec_appe}.

\begin{table*}
\caption[]{
X-ray properties of HRI-detected point sources within and close to 
the optical D25 confines of NGC~300 (see text).
}
\label{table_Hsrc}
\begin{tabular}{lrrrrrrrrr}
\hline
\noalign{\smallskip}
Src.& \multicolumn{2}{c}{R.A.\,(J2000) Dec} & R$_{err}$ & Lik. & Net counts & Ct.\,rate & $F_{\rm x}$ & $L_{\rm x}$ & 
PSPC ID \\
 & ($^{\rm h~~~m~~~s}$) & ($^{\circ}~~~\arcm~~~\arcsec$) & ($\arcsec$) &  & & 
(ks$^{-1}$) & ($\frac{10^{-14}{\rm erg}}{{\rm cm}^{2} {\rm s}}$) & ($\frac{10^{36}{\rm erg}}{{\rm s}}$) & \\
\noalign{\medskip}
(1) & (2) & (3) & (4) & (5) & (6) & (7) & (8) & (9) & (10) \\
\noalign{\smallskip}
\hline
\noalign{\smallskip}
 H6& 00 55 05.65 & -37 35 08.3&  5.8 &   9.2&  12.6(4.4) &  0.3(1) & 1.6(0.6) &  8.6 &    	\\
 H8& 00 55 27.61 & -37 36 53.5&  5.9 &  11.5&  18.3(5.5) &  0.5(1) & 2.4(0.7) & 12.6 & P25 	\\
 H9& 00 55 20.58 & -37 38 48.5&  6.1 &   9.4&  14.7(4.9) &  0.4(1) & 1.9(0.6) & 10.0 &     	\\
H10& 00 54 50.24 & -37 38 49.5&  5.4 &  22.3&  25.8(6.0) &  0.7(2) & 3.3(0.8) & 17.4 & P32 	\\
H11& 00 54 40.65 & -37 40 49.9&  5.4 &  23.2&  35.1(7.1) &  0.9(2) & 4.5(0.9) & 23.7 & P38 	\\
H12& 00 54 45.04 & -37 41 48.5&  6.2 &  11.1&  18.5(5.5) &  0.5(1) & 2.4(0.7) & 12.4 & P41 	\\
H13& 00 55 10.00 & -37 42 15.8&  4.2 & 720.5& 295.0(17.7)&  7.5(5) &37.7(2.3) &199.1 & P42 	\\
H14& 00 55 32.46 & -37 44 07.7&  6.9 &   8.5&  20.0(6.3) &  0.5(2) & 2.6(0.8) & 13.8 &    	\\
H15& 00 54 33.59 & -37 44 48.1&  5.5 &  12.3&  14.7(4.6) &  0.4(1) & 1.9(0.6) & 10.0 & P50	\\
H16& 00 55 10.89 & -37 48 40.5&  4.6 & 179.0& 118.2(11.9)&  3.1(3) &15.5(1.6) & 81.8 & P58	\\
\noalign{\smallskip}
\hline
\end{tabular}
\end{table*}

\begin{figure*}[thpb]
  \resizebox{14cm}{!}{
    \psfig{figure=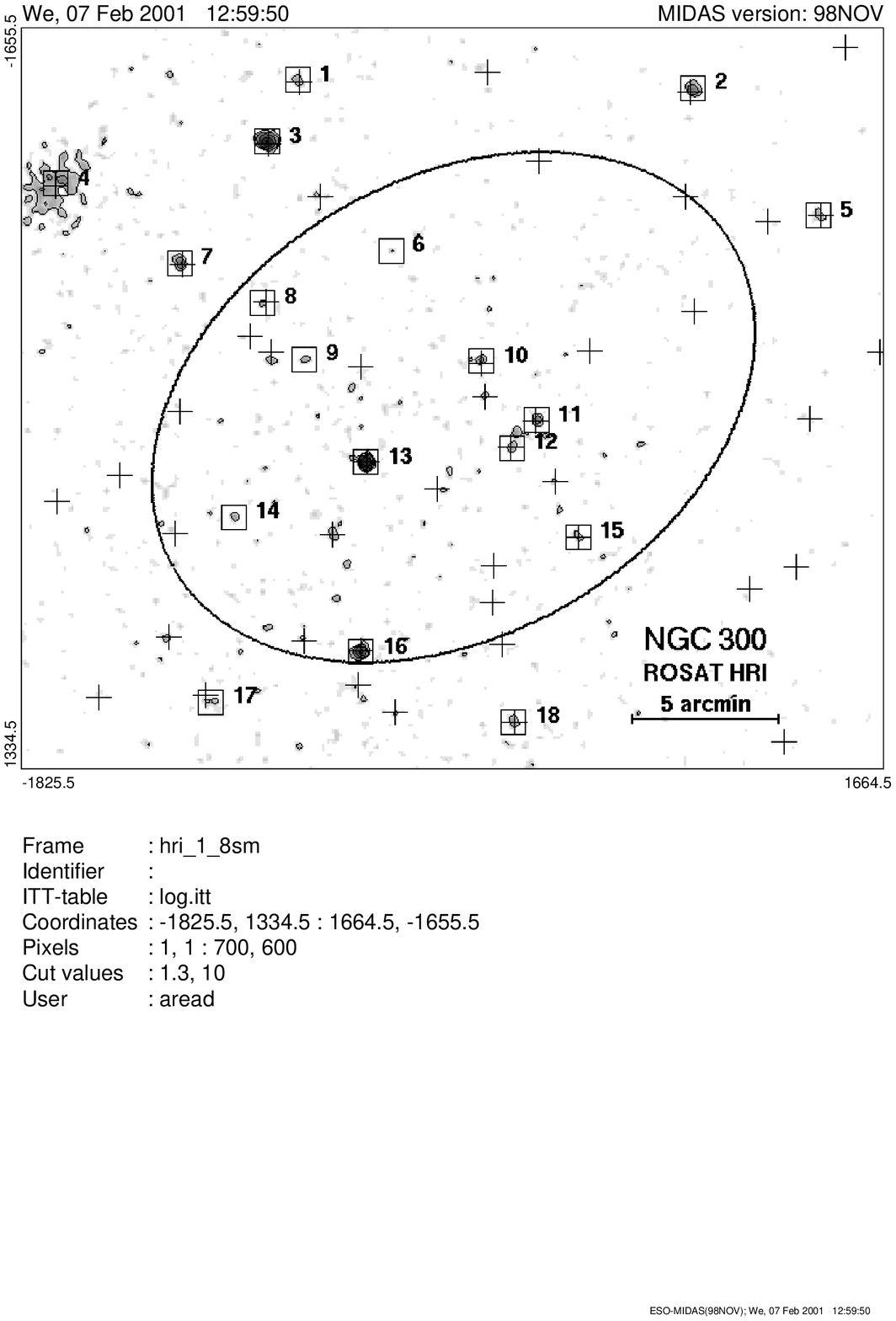,width=14cm,clip=}}
\hfill \parbox[b]{3.5cm}
{\caption{
\Ros\ HRI map of the NGC~300 field (area as in Figs.\,1 \& 3), using HRI
channels 1$-$8). In this non-background-limited case, the contour levels are
at 2, 3, 5, 9 and 15 times a value of $3.1\times10^{-3}$\,cts s$^{-1}$
arcmin$^{-2}$. The D25 ellipse of the galaxy is also marked. The contour
levels HRI Source positions, as given in Table~\ref{table_Hsrc}, are marked by
boxes and numbered on the image. The PSPC source positions (crosses) and the
D25 ellipse of the galaxy are also marked.}}
\end{figure*}

\subsection{Time variability study of NGC~300 point sources}
\label{sec_time}

As mentioned earlier, two separate PSPC observations were carried out, a half year 
apart. The second of these (the larger) consisted of two, quite distinct periods 
of observation, separated by about three weeks. The complete observation was binned 
into these three observation periods; the December 1991 observation, and the two 
halves of the larger observation; the May 1992 and June 1992 observations. 

A maximum likelihood search at the source positions given in
Table~\ref{table_Psrc} was performed for the three observation intervals, the
vignetting and deadtime corrected count rates (and errors) calculated within a cut
radius of $1.5\times$ the PSF FWHM at the source positions.
Table~\ref{table_vari} gives the count rates for the 26 sources for the three
observation intervals plus the probability that the source is variable. 
As can be seen, a few of the sources appear variable, notably source P33. Two further
sources that appear to be variable at greater than the 2$\sigma$ significance level,
are sources P42 ($\sigma=2.3$) and P58 ($\sigma=2.5$).
 
Because of the far fewer photons obtained with the HRI than with the PSPC, no large HRI
time variability study was possible. Also, as we have, for the majority of sources,
insufficient spectral information, no attempt was made at a variability study using
variations between the PSPC and HRI count rates.

\begin{table}
\caption[]{NGC~300 PSPC sources: variability. 
X-ray PSPC count rates of the 26 NGC~300 PSPC sources in the December 1991, 
May 1992 and June 1992 observations (see text). Also given is the probability 
that the source is variable (see text).} 
\label{table_vari}
\begin{tabular}{lrrrr}
\hline
\noalign{\smallskip}
Src. & \multicolumn{3}{c}{PSPC Count rate (ks$^{-1}$)} & Probability  \\
       & Dec.\,'91 & May\,'92 & Jun.\,'92 &         (variation)  \\
\noalign{\medskip}
(1) & (2) & (3) & (4) & (5) \\
\noalign{\smallskip}
\hline
\noalign{\smallskip}
P15&  1.5(0.6) & 1.4(0.5) & 1.2(0.4) & 22\% \\
P17&  1.2(0.5) & 1.5(0.4) & 0.8(0.4) & 44\% \\
P18&  1.4(0.6) & 0.7(0.4) & 1.0(0.4) & 37\% \\
P25&  1.2(0.5) & 2.1(0.5) & 0.9(0.4) & 60\% \\
P26&  0.9(0.5) & 0.5(0.4) & 0.4(0.3) & 35\% \\
P28&  1.5(0.8) & 0.9(0.4) & 0.2(0.3) & 62\% \\
P29&  0.1(0.4) & 0.9(0.4) & 1.4(0.4) & 72\% \\
P30&  1.1(0.7) & 0.9(0.5) & 1.0(0.4) & 18\% \\
P32&  1.1(0.5) & 1.7(0.4) & 1.2(0.4) & 42\% \\
P33&  0.0(0.4) & 7.4(0.8) & 7.5(0.8) &100\% \\ 
P35&  0.5(0.4) & 1.3(0.4) & 1.0(0.4) & 52\% \\
P36&  0.5(0.5) & 0.8(0.4) & 1.2(0.4) & 43\% \\
P38&  3.3(0.8) & 3.1(0.6) & 2.3(0.4) & 50\% \\
P41&  2.3(0.7) & 2.4(0.5) & 2.4(0.5) & 16\% \\
P42& 21.0(1.6) &22.6(1.3) &27.1(1.3) & 80\% \\
P44&  0.7(0.5) & 1.4(0.4) & 1.7(0.4) & 55\% \\
P45&  0.6(0.4) & 0.0(0.4) & 1.4(0.4) & 71\% \\
P48&  0.5(0.5) & 1.9(0.5) & 1.2(0.4) & 65\% \\
P49&  1.7(0.6) & 1.3(0.4) & 1.8(0.5) & 35\% \\
P50&  1.1(0.5) & 0.7(0.4) & 0.8(0.3) & 31\% \\
P51&  0.7(0.6) & 1.2(0.4) & 0.3(0.3) & 62\% \\
P54&  0.9(0.5) & 1.4(0.4) & 0.9(0.3) & 36\% \\
P56&  4.4(0.8) & 2.0(0.5) & 3.4(0.6) & 75\% \\
P57&  0.2(0.3) & 0.7(0.4) & 0.5(0.3) & 47\% \\
P58&  7.5(1.1) & 5.6(0.7) & 9.3(0.8) & 83\% \\
P59&  1.8(0.7) & 2.0(0.5) & 1.7(0.5) & 23\% \\
\noalign{\smallskip}
\hline
\end{tabular}
\end{table}

\section{Results}
\label{sec_resu}

\subsection{Point sources within NGC~300}
\label{sec_res1}

As can be seen in Fig.\,1, 26 PSPC sources are detected within or close to the D25
ellipse of NGC~300, and various X-ray properties of these are given in
Tables~\ref{table_Psrc} and \ref{table_vari}. 10 HRI sources (the majority having
PSPC counterparts) are also detected within the NGC~300 D25 ellipse (Fig.\,3), and
these are listed in Table~\ref{table_Hsrc}. Table~\ref{table_iden} summarises the
following section, giving the best identification of each X-ray source as discussed
here. Firstly (Col.\,3), the corresponding X-ray source from RPS97 (also P00) is
given. Then (Col.\,4) a `B', `F' or `-' is given, indicating whether a bright, a
faint or no DSS2 source is coincident (see Fig.\,1). In confused cases (essentially
within the central NGC~300 regions), a `c' is given. A source identification is
then given (Col.\,5), curved brackets indicating that the identification is not
definite, question marks indicating that it is significantly less so. Square
brackets give the actual names of any particular SNR or \hii\ counterparts (from
BL97 or P00).

Haberl \& Pietsch (1999) found that, in classifying the PSPC-detected X-ray sources
in the LMC, different sources (\eg foreground stars, SSSs and SNRs) show
distinctive hardness ratios on account of their different X-ray spectra. A very
similar situation has also been seen in the SMC and in M33. In Fig.\,4, hardness
ratio 1 (HR1) versus hardness ratio 2 (HR2) is plotted for the 26 sources discussed
above lying within or close to the D25 NGC~300 ellipse. On the plot, lines
delineate the parameter space regions found by Haberl \& Pietsch (1999) to be
populated with distinctly different sources, as follows; A: supersoft X-ray sources
(SSSs), B: foreground stars, C: supernova remnants, D: mixture of all types (though
mainly XRBs and AGN). Also indicated are sources for which we believe we have a
fairly good identification, though an identification {\em not} based on any
hardness ratio arguments, \ie based purely on other observational evidence. It is
worth noting that variations in absorbing column both to and within NGC~300 and M31
will alter the positions of the lines in Fig.\,4 somewhat, but only by a small
amount ($\sim0.1$ in HR) compared to the typical source HR errors. 
We have here
used this hardness ratio information together with the variability studies
presented earlier and observations at other wavelengths in classifying the NGC~300
sources. Much of the information is summarized in Table~\ref{table_iden}, though some
additional explanation and clarification as regards a few of the individual
sources is also given here.

A point to bear in mind is the subject of background AGN. 
An estimate of the number of background sources (predominately AGN and QSOs)
expected at or above our limiting flux level within a sky area the size of the
NGC~300 D25 ellipse ($\sim0.04$\,sq.\,degrees) can be obtained by making use of the
$\log{N}-\log{S}$ function from Hasinger \etal's (1998) deep ROSAT X-ray survey of
the Lockman Hole. This number turns out to be about 4$-$5, though we note that the
nearer their projected position to the centre of NGC~300, the harder they would be
to detect. 

One can also compare the surface density of PSPC and HRI sources inside and outside
of the NGC~300 D25 ellipse, to obtain a perhaps more direct measure of the AGN
contamination within the galaxy. Counting the number of unidentified PSPC sources
outside of the D25 ellipse and scaling this for the areas in question, one obtains
a figure of $\sim$9 for the number of unidentified (possibly background AGN) PSPC
sources expected within a NGC~300 D25 ellipse in this part of the X-ray sky. The
equivalent figure using the HRI data comes out rather lower, more like 3$-$4.
Problems with this method include the fact that all of the unidentified non-D25
sources are unlikely to be AGN, and some of course, may be in reality truly
associated with NGC~300. All this said, we expect the AGN contamination within the
NGC~300 disk to be anything from $\approx4$ to perhaps more than double this value.
The source identification column (Col.\,5) of Table~\ref{table_iden} includes `AGN'
entries, where we are fairly sure that the source is a background AGN, but also
`acc.' entries, where the source appears to be accreting in nature, though whether
it is background or related to NGC~300 is difficult to say.

\begin{figure}
\unitlength1.0cm 
\begin{picture}(8.8,6.4)
\psfig{figure=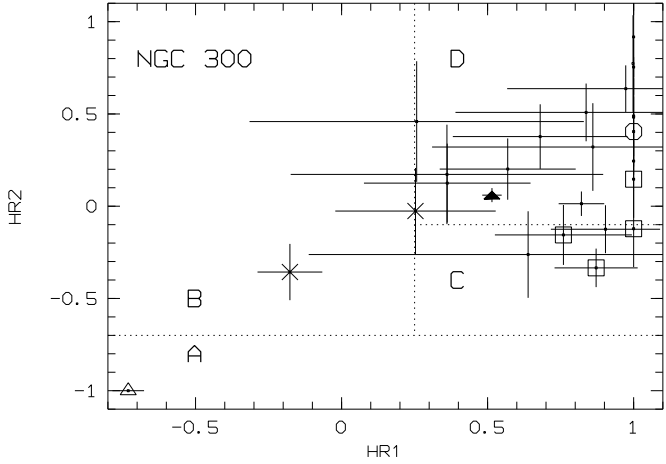,width=8.8cm,clip=}
\end{picture} 
\hfill \parbox[b]{8.8cm} 
{\caption{
Hardness ratio classification of the PSPC-detected sources within NGC~300. 
Areas A$-$D of the parameter space are populated 
by different sources; A: supersoft X-ray sources (SSSs), B: foreground stars, 
C: supernova remnants, D: mixture of all types (though mainly XRBs and AGN). 
Identified stars are marked with X's, SNRs with boxes, SSS candidates with 
triangles, XRB candidates with filled triangles and AGN candidates with 
hexagons. 
}}
\end{figure}

All the sources detected by RPS97 
(and used in the P00 study) are detected. All 12 of the newly detected NGC~300 
PSPC sources have a net counts value less than 50 (except for P59, which, though 
significantly visible in the RPS97 study, lay outside their area of interest, and 
was hence not included in their list). All 14 RPS97 sources re-detected here 
have a net counts value greater than 50, except for sources P29 and P30, the 
least significant of the sources listed in RPS97/P00. (Note RPS97 source 14 
corresponds to non-D25 source P24). 

\begin{table}
\caption[]{
The best identifications for each NGC~300 
X-ray source. Firstly, the corresponding X-ray source from RPS97 (also P00) is 
given. 
A `B', `F' or `-' is then given, indicating whether a bright, a faint or no DSS2 source is 
coincident. In confused cases (essentially within the central 
NGC~300 regions), a `c' is given. A source identification is lastly given, curved brackets indicating 
that the identification is not definite, question marks 
indicating that it is significantly less so. Square brackets give the actual names of 
any particular counterparts (from BL97 or P00). 
}
\label{table_iden}
\begin{tabular}{ccccr}
\hline
\noalign{\smallskip}
PSPC & HRI & \multicolumn{3}{c}{Identifications} \\
Src. & Src.& \multicolumn{2}{c}{(RPS97) (DSS2)}   &  (type/[name])          \\
\noalign{\medskip}
(1) & (2) & (3) & (4) & (5) \\
\noalign{\smallskip}
\hline
\noalign{\smallskip}
P15&     & 3 & - & (Star?) 		\\
P17&     &   & F & (AGN)   		\\
P18&     &   & - & (Star?) 		\\
   &  H6 &   & F & (?)        	\\
P25&  H8 &13 & F & (Acc.?) 		\\
P26&     &   & F & (Acc.?) 		\\
P28&     &   & F & (Acc.?) 		\\
P29&     & 1 & c & SNR[BL97-S6] 		\\
P30&     &12 & F & (Acc.?) 		\\
   &  H9 &   & c & (?) 		\\
P32& H10 & 6 & c & (XRB) 		\\
P33&     & 9 & c & SSS 		\\
P35&     &   & c & \hii[BL97-H13] (SNR) 	\\
P36&     &   & F & (SNR) 		\\
P38& H11 & 4 & c & SNR[BL97-S10] 	\\
P41& H12 & 5 & c & (SNR) 		\\
P42& H13 & 7 & c &  BHXRB 		\\
P44&     & 2 & c & (XRB) ([P00-SNR3]?)	\\
P45&     &   & c & Star 		\\
   & H14 &   & - & (?) 		\\
P48&     &15 & B & (Star) 		\\
P49&     &10 & c & SNR[BL97-S26] 	\\
P50& H15 &   & F & (?) 		\\
P51&     &   & F & (Acc.?) 		\\
P54&     &   & - & (Acc.?) 		\\
P56&     &11 & F & (Star) ([P00-SNR15]?)		\\
P57&     &   & - & (Acc.?) 		\\
P58& H16 & 8 & B & (AGN) 		\\
P59&     &   & F & (AGN) 		\\
\noalign{\smallskip}
\hline
\end{tabular}
\end{table}

The brightest source, P42, lies within the inner spiral arms of NGC~300. No SNR, giant
\halpha\ or \hii\ regions are seen coincident (BL97), and it is likely an accreting
binary. Note that the Eddington limit for a $1M_\odot$ compact object is $1.3\times
10^{38}$\,erg~s$^{-1}$, so source P42's X-ray emission may suggest the prescence of a
more massive ($\sim5M_\odot$) black hole X-ray binary.
P33 is both extremely soft (note the hardness ratios given in Table~\ref{table_Psrc} and
the soft- and hard-band images in Fig.\,2) and extremely variable; it is not detected at
all in the December 1991 observation, but is seen to be very bright in the May/June 1992
observation. Also the source is not detected at all in any of the HRI observations (see
Fig.\,3). No counterparts are seen, and it appears to be another example of a supersoft
source (SSS), extremely soft X-ray sources, quite often seen to be variable (see Kahabka
\& van den Heuvel 1997 for a review).
P38 (H11) lies coincident with a SNR (S10; BL97), also confirmed by optical spectroscopy
(DDB2; D'Odorico \& Dopita 1983). Close by, P41 (H12) is similarly HR2-soft and is
non-variable. Two bright \halpha\ knots are visible, and this source may also be a SNR, as
also suggested by P00.
P49 lies 6.6\arcs\ offset from SNR S26 (BL97), also confirmed by optical spectroscopy
(DDB5; D'Odorico \& Dopita 1983). Although not formally detected in the HRI data, some
HRI emission from this position is evident (Fig.\,3).
P32 (H10) has no nearby counterparts, and is likely to be an accreting binary.
Similarly, P44 lies coincident with bright \halpha\ knots, and P00
suggest that it may be associated with a possibly radio-detected SNR (their SNR3). Given
its hard X-ray HR2 however, it is more likely to be accreting in nature. Interestingly,
the X-ray feature visible just to the south-east of P44 is coincident with a SNR (S11;
BL97). It may be worth noting here that, even with the greater sensitivity of the
present work over RPS97, and the larger source list, no further X-ray counterparts to
the new candidate SNRs of P00 are detected (apart from the ones listed in P00).
Sources P29 and P35 are similarly bright and hard. P29 lies less than 1\arcs\ from SNR
S6 (BL97), and the hint of variability and the spectral hardness may indicate that some
emission from a more compact harder component, perhaps at the centre of the remnant, is
contributing. P35, coincident with a \hii\ region (No.\,13; BL97), is likely also
SNR-related. Lastly, the hardness of P45 points to it being a (very variable) foreground
star.

To the north, the hard P17 has a DSS2 optical counterpart, and is likely a background
AGN, while sources P15 and P18, given their hardness ratos, are more likely stellar in
origin. To the south, P58 (H16), and P59 are highly likely to be associated with distant
AGN. Though P00 discuss P56 (or rather RPS97 source 11) as a candidate SNR (their
SNR15), given P56's softness, and its position at the very edge of the NGC~300 disk, the
likely counterpart is probably a foreground star. Note that HRI emission is evident at
the positions of P59 and P56. Sources P51, P54 and P57, given their hardness, are
probably accreting objects, though whether they are foreground objects, objects within
NGC~300, or distant AGN is difficult to say. To the east, P48 lies exactly coincident
with a very bright DSS2 point source (very apparent in Fig.\,1), and is likely a
foreground star. P36 has a very faint optical counterpart and is more likely
SNR-related. As regards P25 (H8), P28 and P30, faint DSS2 counterparts do exist in each
case, and only conclusions similar to those drawn as regards the P51/P54/P57 triple can
be drawn here.

The formal HRI detection algorithm picks up a further three sources within the D25
ellipse not formally detected in the PSPC, namely H6, H9 \& H14. These though are
detected with the lowest values of existence likelihood (less than 10), and cannot be
solidly believed without additional evidence. No SNRs, \hii\ regions or significant DSS2
sources are visible at the positions of these three.

\subsection{Other point sources}
\label{sec_res2}

Although we here concentrate on the point source population within NGC~300, 
a few of the surrounding sources are interesting, and it is worth 
referring to Figs.\,1 \& 3 and to Tables~\ref{table_PsrcA} and 
\ref{table_HsrcA} in Appendix~\ref{sec_appe}. 

The non-variable P14 (H3) is associated with the prominent bright star, 
HD5403 (visible to the north-east of NGC~300 in Fig.\,1), 
and has been used to improve both the X-ray PSPC and HRI positions. 

North-east and east of P14 lies the ($z=0.055$) galaxy cluster Abell~S0102 (Abell
\etal\ 1989), the central galaxies of which themselves show some extremely interesting
X-ray and radio properties (Read \etal\ 2001). Cappi \etal\ (1998) have recently shown
that part of the Abell~S0102 cluster is in fact a separate, more distant ($z=0.165$) 
cluster, CL~0053-37, and this cluster is entirely coincident with P16 (H4), the highly 
extended (likelihood of extent being [PSPC] $\sim1500$, [HRI] $\sim83$), 
bright source to the far north-east of Fig.\,1. Several distant galaxy-like sources 
are visible in the optical associated with P16 (Fig.\,1). 
Again, no evidence for PSPC or HRI variability is observed. 

P37 appears marginally extended (likelihood of extent, 9.8), perhaps due to its
emission being from two bright optical sources in the vicinity, one of which is also an
NRAO VLA radio source (NVSS~J005353-374020; Condon \etal\ 1998). Similarly, P47
(again of marginal extent, the extent likelihood being 9.9) appears likely associated
with QSO~005342.1-375947 (Iovino \etal\ 1996), P53 with the radio source
MRC~0051-380 (Condon \etal\ 1998), and P63 with the galaxy B005241.34-380658.4
(Maddox \etal\ 1990).


\section{Discussion}
\label{sec_disc}

\subsection{Emission components within NGC~300}
\label{sec_disc1}

We have attempted to identify above as many of the X-ray sources within NGC~300 as
possible. Once we eliminate from the source list those sources we are fairly sure do
not belong to NGC~300, namely (see Table~\ref{table_iden}) sources P17, P45, P48,
P56, P58 \& P59, then one arrives at a total (0.1$-$2.4\,keV) X-ray luminosity of
point sources within NGC~300 of $4.6\times10^{38}$\,erg s$^{-1}$ (assuming a distance
of 2.1\,Mpc). 
Note that the number of believed background AGN excluded from the list is 
consistent with the number expected statistically from deep ROSAT surveys.  
Note though that we have been rather conservative in removing the sources $-$ a number 
of the `acc.' sources may well be unrelated to NGC~300 also.

Comparing the PSPC broad-band total counts within the NGC~300 D25 ellipse with the
counts detected in (all) the D25 sources, one detects a small amount of residual
emission, presumably due to unresolved point sources and/or genuine hot gas. In order
to calculate this residual non-source NGC~300 flux, one needs to take the background
into account. Variations in the background outside of NGC~300 and of source
regions however, are comparable with the level of the residual NGC~300 emission itself,
and so only a very approximate value of $\sim300\pm200$ residual NGC~300 PSPC counts can
be estimated. This small number of counts, together with the large uncertainties in the
background outside of NGC~300 made any spectral fitting of this residual emission
rather impossible. Assuming a 0.25\,keV thermal bremsstrahlung model, typical of
relatively inactive normal spiral galaxies (\eg RPS97), with a hydrogen
column density of $N_{\rm H} = 3.6 \times 10^{20}$~cm$^{-2}$ (\ie no intrinsic
NGC~300 absorption), then we arrive at a (0.1$-$2.4\,keV) flux of
$2.2\times10^{-13}$\,erg cm$^{-2}$ s$^{-1}$, and a luminosity ($D=2.1$\,Mpc) of
$1.2\times10^{38}$\,erg s$^{-1}$. This gives rise to a total (0.1$-$2.4\,keV) X-ray
luminosity, \ie sources plus residual (diffuse gas and unresolved sources) emission,
of $5.8\times10^{38}$\,erg s$^{-1}$, the residual emission fraction (the fraction of
the luminosity due to residual emission) being $\sim20$\%. 

\subsection{The X-ray source luminosity distribution in NGC~300}
\label{sec_disc2}

Figure\,5 shows the luminosity distribution of the X-ray sources in NGC~300, as a function
of the mass of neutral hydrogen. Also shown, for comparison, are the corresponding
distributions for seven other well-known nearby spirals galaxies (strong nuclear 
sources having been removed), this adapted from
Fig.\,14 of Wang \etal\ (1999), the values for M83 having been updated (Immler \etal\
1999). All values for the neutral hydrogen mass are taken from Tully (1988).

\begin{figure}
\unitlength1.0cm 
\begin{picture}(8.8,7.1)
\psfig{figure=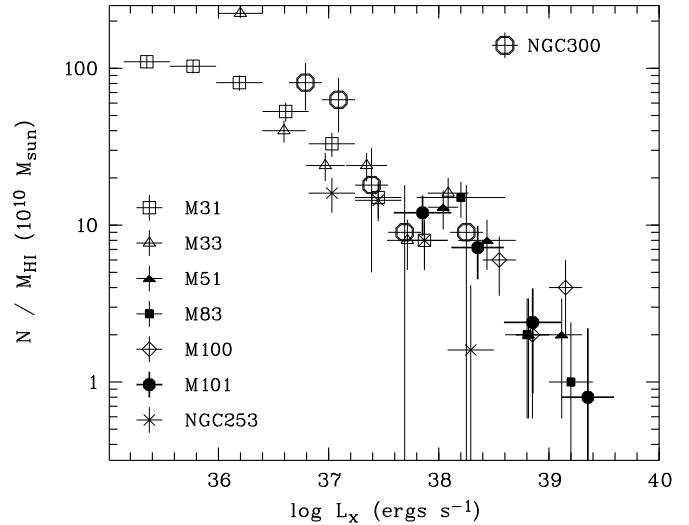,width=8.8cm,clip=}
\end{picture} 
\hfill \parbox[b]{8.8cm} 
{\caption{
X-ray Luminosity distribution of sources within NGC~300 (large hexagons), as a function
of neutral hydrogen mass, compared with seven other well-known nearby spiral galaxies
(adapted from Wang \etal\ 1999; see text).
}}
\end{figure}

Though in the case of NGC~300, we are dealing with rather low number statistics when
compared with the other systems (hence the larger errors), NGC~300's X-ray source
luminosity distribution does not appear significantly different from the others, and all
the galaxies appear to have very similar \hi-normalised luminosity distributions. The
suggestion of an increase in NGC~300's distribution at the low-luminosity end may be
attributed to a small underestimation of the \hi\ mass of NGC~300 in Tully (1988)
compared with more recent studies, \eg Puche \& Carignan (1991) (the Tully (1988) value
has been used in Fig.\,5 for the sake of consistency with the other systems). 
It is 
certainly true that the perhaps `relative steepness' of NGC~300's source luminosity
distribution can also be brought into line with the other systems, if 2 or 3 of the
questionable `acc.' sources (see Table~\ref{table_iden}) are removed. We have been here 
slightly conservative, in removing only sources we are fairly sure do not belong to
NGC~300 $-$ it may well be the case, and Fig.\,5 seems to bear this out, that a couple
of the low luminosity sources may well not belong to NGC~300. Finally note that,
although no very bright sources are seen in NGC~300, none are also seen in the other
normal, quiescent galaxies M31 and M33. NGC~300 therefore, in terms of the luminosity
distribution of its X-ray sources, appears in no way unusual.

\subsection{NGC~300 as a member of the Sculptor group}
\label{sec_disc3}

The Sculptor group is probably the nearest small group of galaxies to our own Local
Group, and the individual members have been studied in great detail. Kinematical
studies (\eg Puche \& Carignan 1988) have established the group to be made
up of five major members (NGC~55, NGC~247, NGC~253, NGC~300 and NGC~7793). Other
smaller galaxies, such as NGC~24 and NGC~45, were found to be more distant, and
although dwarf galaxies do exist within Sculptor (\eg Lausten \etal\ 1977),
they contribute essentially nothing to the group emission or dynamics.

The X-ray properties of the five members have been presented several times in the past
(\eg Schlegel \etal\ 1997; RPS97; Vogler \& Pietsch 1999; Pietsch \etal\ 2000), and in
discussing the \Ros\ PSPC observations of NGC~7793 (Read \& Pietsch 1999, hereafter
RP99), a comparison of the Sculptor galaxies' X-ray properties was presented, and it is
well worth referring to RP99 here. A similar table to Table~5 of RP99 is given in
Table~\ref{table_scul}. Various physical, X-ray and other multi-wavelength properties
for the five prominent Sculptor members are given. All X-ray information for NGC~300 is
taken from the present paper, while that for the remaining systems is taken from RPS97,
apart from NGC~7793 (RP99). 
Given in the last column (12) is the
$\log{L_{X}/M_{\mbox{\small HI}} }$ ratio, the neutral hydrogen mass values taken from
Tully (1988).

\begin{table*}
\caption[]{
Properties of the Sculptor group galaxies (see text and RP99 for details). 
All X-ray information for NGC~300 is taken from the present paper. All X-ray information for the
remaining systems is taken from RPS97, apart from NGC~7793 (RP99). Neutral hydrogen mass values are 
taken from Tully (1988).}
\label{table_scul}
\begin{tabular}{lrrrrrrrrrrr}
\hline
\noalign{\smallskip}
Galaxy & Type & Diam. & Axis & Dist. & \multicolumn{3}{c}{Log luminosity (erg s$^{-1}$)} &
 \multicolumn{3}{c}{Luminosity ratios ($10^{-4}$)} & $\log{L_{X}/M_{\mbox{\small H{\rm I}}} }$\\
       &      & (kpc) & ratio& (Mpc) & $L_{B}$ & $L_{FIR}$ & $L_{X}$ & $L_{FIR}/L_{B}$ & $L_{X}/L_{B}$
 & $L_{X}/L_{FIR}$  & (erg s$^{-1}$/$M_{\odot}$) \\
(1) & (2) & (3) & (4) & (5) & (6) & (7) & (8) & (9) & (10) & (11) & (12) \\
\noalign{\smallskip}
\hline
\noalign{\smallskip}
 NGC~55  & SBS9 & 12.3 & 5.8 & 1.3 & 43.02 & 41.94 & 38.91 & 830 & 0.78 & 9.3 & 29.86 \\
 NGC~247 & SXS7 & 13.1 & 3.1 & 2.1 & 42.96 & 41.48 & 38.68 & 330 & 0.52 &16.0 & 29.78 \\
 NGC~253 & SXS5 & 24.0 & 4.1 & 3.0 & 43.78 & 43.74 & 40.04 &9100 & 1.80 & 2.0 & 30.26 \\
 NGC~300 & SAS7 & 13.3 & 1.4 & 1.2 & 43.01 & 41.92 & 38.76 & 810 & 0.57 & 6.9 & 29.86 \\ 
 NGC~7793& SAS7 &  9.1 & 1.5 & 3.4 & 43.01 & 42.23 & 38.78 &1700 & 0.59 & 3.6 & 30.08 \\ 

\hline
\end{tabular}
\end{table*}

What is interesting as regards NGC~300 is that now, the galaxy appears more `normal' than
previously reported. In RP99, the NGC~300 data reported, analysed in
RPS97, assumed a distance of 1.2\,Mpc (Tully 1988). Now we have assumed a
distance, as described in the introduction, of 2.1\,Mpc, based on several authors' more
recent work. Note that, correcting for the different assumed distances, the
luminosity for NGC~300 quoted here and that in the RPS97 survey, 
where only a part of the NGC~300 PSPC data was analysed (and then only in a
semi-automatic sense), agree rather well.

The most famous Sculptor member is probably NGC~253, a large X-ray and far-infrared
bright starburst galaxy, while the remaining four systems are all very similar;
physically small, normal (\ie non-starburst), late-type spirals (NGC~253 is also a
late-type spiral). The four normal Sculptor systems are also extremely similar in terms
of their multiwavelength properties. The narrow range in especially $L_{B}$, but also
$L_{X}$, evident in Table~\ref{table_scul} (excluding NGC~253) is very striking (the
larger range in $L_{FIR}$ observed is partly due to properties of NGC~7793 - see RP99).

Though this in itself is interesting, \ie that the four normal Sculptor galaxies are
extremely similar, and are therefore good examples of prototypical `normal' galaxies,
what is perhaps remarkable about NGC~300 is that it is the most `normal' of the four. It is
neither the brightest nor the dimmest galaxy in either the optical, the far-infrared or
the X-ray band. Similarly, it has neither the largest nor the smallest value of
$L_{FIR}/L_{B}$, $L_{X}/L_{B}$ or $L_{X}/L_{FIR}$ flux ratio. Furthermore, \hi\ studies
and mass modelling of the Sculptor group galaxies (\eg Puche \& Carignan 1991) have
shown that NGC~300 lies midway in every parameter space, \eg mass-to-light ratio
$M/L_{B}$, \hi\ mass \etc\ Note also NGC~300's midway 
$\log{L_{X}/M_{\mbox{\small HI}} }$ ratio. 

As seen in Table~\ref{table_scul}, the other three normal Sculptor galaxies are unusual
in some way; NGC~55 is X-ray bright, NGC~7793 is very far-infrared bright and NGC~247 is
very far-infrared dim. NGC~300 however, is wholly unremarkable, both in terms of how it
compares with its Sculptor neighbours and, as seen in Sect.~\ref{sec_disc2}, in how its
X-ray source luminosity distribution compares with other nearby spiral galaxies. It is
therefore perhaps one of the finest examples of a typical quiescent 
normal late-type spiral galaxy.

\section{Summary}
\label{sec_summ}

We have analysed all the \Ros\ PSPC and HRI data from a field centred on the nearby
face-on Sculptor galaxy NGC~300. 29 PSPC and HRI sources are detected within the 
optical confines of the galaxy. Many of the sources appear to be variable and we attempt to
classify and identify the sources based on their temporal and spectral properties and 
using other multi-wavelength observations. In
addition to point source emission, some evidence for unresolved residual emission is
detected within NGC~300. Our findings with regard to the observed point-source and
unresolved emission can be summarized as follows:

1. 26 PSPC sources are detected within the D25 ellipse of NGC~300, as are 10 HRI 
sources (all but the very dimmest having PSPC counterparts), and we present full 
source lists for these, detailing their X-ray properties. 

2. The brightest source ($L_{X} (0.1$-$2.4\,keV) = 2.2\times10^{38}$\,erg s$^{-1}$) is 
likely a BHXRB, while a highly variable source nearby is seen to be a supersoft source. 
Other bright sources coincident with known SNRs and \hii\ regions are discussed, as are 
foreground stars and background AGN. 

3. The present analysis attains a greater sensitivity than previous studies, and 
the number of believable X-ray sources within NGC~300 has been increased almost 
twofold. That said, no further X-ray counterparts to new candidate SNRs have been 
detected. 

4. Residual X-ray emission is observed within NGC~300, due probably to unresolved 
sources and genuine diffuse gas. This emission ($L_{X} = 1.2\times10^{38}$\,erg s$^{-1}$)
accounts for approximately 20\% of the total X-ray luminosity of NGC~300 
($5.8\times10^{38}$\,erg s$^{-1}$). 

5. Detailed PSPC and HRI source lists are also presented for sources detected 
outside of NGC~300, and we briefly discuss a number of them, notably the galaxy 
cluster CL\,0053-37. 

6. The X-ray source luminosity distribution of NGC~300 is in no way unusual when 
compared with other nearby spiral galaxies. The observed lack of highly luminous 
X-ray sources is entirely consistent with other normal, quiescent spirals. 

7. In terms of how the X-ray properties of NGC~300 compare with those of 
its neighbours in the Sculptor galaxy group, NGC~300 appears to be the most 
unremarkable of all the normal group members. It shows no unusual X-ray properties, 
nor oddities in multi-wavelength luminosities or ratios. It may the best example 
of a typical normal quiescent late-type spiral galaxy.

\begin{acknowledgements}

We thank our colleagues from the MPE ROSAT group for their support, and 
we are especially grateful to Rodrigo Supper for reading the manuscript, 
and to the referee for comments which have greatly improved the paper.
The ROSAT
project is supported by the German Bundesministerium f\"ur Bildung und
Forschung (BMBF) and the Max-Planck-Gesellschaft (MPG).
Based on photographic data obtained using The UK Schmidt Telescope.
The UK Schmidt Telescope was operated by the Royal Observatory
Edinburgh, with funding from the UK Science and Engineering Research
Council, until 1988 June, and thereafter by the Anglo-Australian
Observatory.  Original plate material is copyright (c) the Royal
Observatory Edinburgh and the Anglo-Australian Observatory.  The
plates were processed into the present compressed digital form with
their permission.  The Digitized Sky Survey was produced at the Space
Telescope Science Institute under US Government grant NAG W-2166.

\end{acknowledgements}

\appendix

\section{ROSAT PSPC and HRI sources outside of NGC~300}
\label{sec_appe}

Here are tabulated the X-ray properties of the PSPC and HRI sources detected outside of the
NGC~300 D25 ellipse.

Table~\ref{table_PsrcA} lists the PSPC values, as follows: source number (Col.\,1), corrected
right ascension and declination (Cols.\,2,\,3), error on the source position (Col.\,4, including a
3\farcs9 systematic attitude solution error), likelihood of existence (Col.\,5), net broad band
counts and error (Col.\,6), and count rates and errors after applying deadtime and vignetting
corrections (Col.\,7), hardness ratios HR1 and HR2 (Cols.\,8 \& 9; see Sect.~\ref{sec_obse}).
Finally, the 0.1$-$2.4\,keV flux and X-ray luminosity, assuming a 5\,keV thermal bremsstrahlung
model and a source distance of 2.1\,Mpc (\ie to NGC~300, which is here unlikely to be valid,
hence the bracketed values), are given in Cols.\,10 \& 11.

Table~\ref{table_HsrcA} lists the HRI values in the same way as Table~\ref{table_PsrcA}, except
that no hardness ratios are given, hence the 0.1$-$2.4\,keV flux and the (again bracketed) X-ray
luminosity are given in Cols.\,8 \& 9. A final identification column is given (Col.\,10), listing
which PSPC source (if any) is a likely counterpart, whether any DSS2 candidates (as for
Table~\ref{table_iden}), and the names of any known counterparts (see Sect.~\ref{sec_res2}).

\begin{table*}
\caption[]{X-ray properties of PSPC-detected point sources outside the 
optical confines of NGC~300 (see text).
Tabulated fluxes and luminosities 
assume a 5\,keV thermal bremsstrahlung model, a hydrogen column
density of $N_{\rm H} = 3.6 \times 10^{20}$~cm$^{-2}$ and an NGC~300
distance of 2.1\,Mpc (luminosities are given in brackets, as the sources are 
unlikely to be associated with NGC~300)}
\label{table_PsrcA}
\begin{tabular}{lrrrrrrrrrr}
\hline
\noalign{\smallskip}
Src.& \multicolumn{2}{c}{R.A.\,(J2000) Dec} & R$_{err}$ & Lik. & Net counts & Ct.\,rate &
\multicolumn{2}{c}{Hardness ratios} & $F_{\rm x}$ & $L_{\rm x}$ \\
 & ($^{\rm h~~~m~~~s}$) & ($^{\circ}~~~\arcm~~~\arcsec$) & ($\arcsec$) &  & &
(ks$^{-1}$) & (HR1) & (HR2) & ($\frac{10^{-14}{\rm erg}}{{\rm cm}^{2} {\rm s}}$) & 
($\frac{10^{36}{\rm erg}}{{\rm s}}$) \\
\noalign{\medskip}
(1) & (2) & (3) & (4) & (5) & (6) & (7) & (8) & (9) & (10) & (11) \\
\noalign{\smallskip}
\hline
\noalign{\smallskip}
 P8& 00 53 48.25 & -37 28 10.5& 7.5& 648.8& 433.2(24.0)&11.5(0.6)& 0.81(05)& 0.18(06)& 20.2(1.1)&(106.6) \\
 P9& 00 54 49.43 & -37 29 02.3&16.2&  22.5&  48.9(10.8)& 1.2(0.3)& 0.10(34)& 0.42(22)&  2.1(0.5)&( 11.3) \\
P10& 00 55 21.53 & -37 29 21.1& 8.3& 149.4& 183.5(17.2)& 4.6(0.4)& 0.17(12)& 0.15(10)&  8.1(0.8)&( 43.0) \\
P11& 00 54 14.70 & -37 29 42.5& 7.1& 407.4& 292.1(20.2)& 7.4(0.5)& 0.68(07)& 0.11(07)& 13.0(0.9)&( 68.7) \\
P14& 00 55 26.80 & -37 31 26.5& 4.8&1966.9&1148.6(36.5)&28.5(0.9)&-0.20(03)&-0.27(05)& 50.2(1.6)&(264.8) \\
P16& 00 56 03.21 & -37 32 49.9& 6.7&1388.7&2779.0(60.9)&73.4(1.6)& 0.74(02)& 0.28(02)&129.2(2.8)&(681.7) \\
P20& 00 53 52.41 & -37 33 51.3&10.9& 102.5& 129.9(15.3)& 3.3(0.4)& 0.33(16)&-0.06(12)&  5.8(0.7)&( 30.5) \\
P21& 00 54 01.39 & -37 34 05.2&21.7&  26.6&  44.1(11.7)& 1.1(0.3)& 1.00(00)& 0.27(20)&  1.9(0.5)&( 10.2) \\
P24& 00 55 41.63 & -37 35 32.0& 5.8& 458.5& 329.6(20.5)& 8.2(0.5)& 0.59(07)&-0.17(07)& 14.3(0.9)&( 75.7) \\
P31& 00 53 42.34 & -37 38 28.3&16.9&  28.7&  40.8(10.4)& 1.0(0.3)& 1.00(00)& 0.42(18)&  1.8(0.5)&(  9.6) \\
P37& 00 53 54.31 & -37 40 44.7&27.6&  24.7&  90.9(17.9)& 2.2(0.4)& 0.38(24)& 0.12(17)&  3.9(0.8)&( 20.8) \\
P43& 00 55 52.37 & -37 42 40.2&24.9&  12.8&  44.5(11.4)& 1.1(0.3)&-0.21(26)& 0.04(32)&  1.9(0.5)&( 10.2) \\
P47& 00 56 03.16 & -37 43 30.9&24.3&  62.4&  68.9(12.3)& 1.8(0.3)& 0.56(17)& 0.19(15)&  3.1(0.6)&( 16.3) \\
P52& 00 53 56.52 & -37 45 45.3&31.1&  10.0&  17.2(9.6) & 0.4(0.2)& 1.00(00)& 0.72(26)&  0.8(0.4)&(  4.0) \\
P53& 00 54 04.39 & -37 46 30.3&18.9&  15.6&  38.8(10.9)& 0.9(0.3)&-0.33(23)& 0.51(28)&  1.7(0.5)&(  8.8) \\
P55& 00 55 44.10 & -37 48 07.3&13.4&  32.6&  46.6(10.6)& 1.2(0.3)& 1.00(00)& 0.06(21)&  2.0(0.5)&( 10.8) \\
P61& 00 55 37.70 & -37 50 05.4&21.9&  12.4&  33.9(10.7)& 0.8(0.3)& 1.00(00)& 0.44(30)&  1.5(0.5)&(  7.9) \\
P62& 00 55 56.15 & -37 50 09.5&14.7&  79.7& 142.1(17.3)& 3.7(0.5)& 0.64(13)& 0.26(11)&  6.5(0.8)&( 34.1) \\
P63& 00 55 05.17 & -37 50 39.0&18.9&  10.7&  28.1(9.3 )& 0.7(0.2)& 0.38(57)& 0.12(34)&  1.2(0.4)&(  6.3) \\
P65& 00 54 44.75 & -37 51 01.2& 8.2& 137.3& 121.1(13.8)& 2.9(0.3)& 0.58(16)& 0.02(12)&  5.2(0.6)&( 27.3) \\
P66& 00 53 58.59 & -37 51 38.7&20.3&  28.4&  53.7(11.7)& 1.4(0.3)& 0.61(35)& 0.58(19)&  2.4(0.5)&( 12.8) \\
\noalign{\smallskip}
\hline
\end{tabular}
\end{table*}

\begin{table*}
\caption[]{X-ray properties of HRI-detected point sources outside  
the optical D25 confines of NGC~300 (see text).
Tabulated fluxes and luminosities 
assume a 5\,keV thermal bremsstrahlung model, a hydrogen column
density of $N_{\rm H} = 3.6 \times 10^{20}$~cm$^{-2}$ and an NGC~300 
distance of 2.1\,Mpc (luminosities are given in brackets, as the sources are 
unlikely to be associated with NGC~300)}
\label{table_HsrcA}
\begin{tabular}{lrrrrrrrrrr}
\hline
\noalign{\smallskip}
Src.& \multicolumn{2}{c}{R.A.\,(J2000) Dec} & R$_{err}$ & Lik. & Net counts & Ct.\,rate & $F_{\rm x}$ & $L_{\rm x}$ & 
\multicolumn{2}{c}{Ident.} \\
 & ($^{\rm h~~~m~~~s}$) & ($^{\circ}~~~\arcm~~~\arcsec$) & ($\arcsec$) &  & & 
(ks$^{-1}$) & ($\frac{10^{-14}{\rm erg}}{{\rm cm}^{2} {\rm s}}$) & ($\frac{10^{36}{\rm erg}}{{\rm s}}$) & & \\
\noalign{\medskip}
(1) & (2) & (3) & (4) & (5) & (6) & (7) & (8) & (9) & \multicolumn{2}{c}{(10)} \\
\noalign{\smallskip}
\hline
\noalign{\smallskip}
 H1& 00 55 21.48 & -37 29 20.6&  8.1 &  11.4&  37.5(9.7) &  1.0(0.3) & 5.1(1.3) &( 27.0) & P10& F 	\\
 H2& 00 54 14.09 & -37 29 38.8&  6.4 &  56.2&  96.7(12.9)&  2.7(0.4) &13.3(1.8) &( 70.3) & P11& F	\\
 H3& 00 55 26.80 & -37 31 26.0&  4.8 & 410.3& 290.5(18.8)&  7.8(0.5) &39.2(2.5) &(206.7) & P14& HD5403      \\
 H4& 00 56 02.91 & -37 32 49.3&  8.0 &  80.3& 475.9(33.3)& 13.4(0.9) &67.0(4.7) &(353.7) & P16& CL0053-37   \\
 H5& 00 53 52.66 & -37 33 54.6&  8.2 &  15.4&  47.6(10.7)&  1.3(0.3) & 6.6(1.5) &( 34.7) & P20& (AGN)	\\
 H7& 00 55 41.72 & -37 35 32.9&  5.8 &  46.3&  63.6(10.0)&  1.7(0.3) & 8.5(1.3) &( 44.9) & P24& F   \\
H17& 00 55 36.77 & -37 50 23.6&  9.6 &   8.3&  34.7(10.1)&  0.9(0.3) & 4.7(1.4) &( 24.8) & P61& -	\\
H18& 00 54 44.66 & -37 51 03.8&  6.6 &  15.3&  31.0(7.6 )&  0.8(0.2) & 4.1(1.0) &( 21.7) & P65& F   \\
\noalign{\smallskip}
\hline
\end{tabular}
\end{table*}

\end{document}